\documentclass[12pt]{article}
\usepackage{amsmath}
\usepackage{amssymb}
\usepackage{epsfig}
\usepackage{fullpage}

\def \BE {\begin{equation}}
\def \EE {\end{equation}}
\def \BEA {\begin{eqnarray}}
\def \EEA {\end{eqnarray}}

\begin{document}

\title{Gravity surface wave turbulence in a laboratory flume}
\author{Petr Denissenko$^1$, Sergei Lukaschuk$^1$ and Sergey Nazarenko$^2$
\\ \ \\
{\small $^1$ Fluid Dynamics Laboratory, the University of Hull,
Hull, HU6 7RX, UK
}\\
$^2$ {\small
Mathematics Institute, The University of Warwick,  Coventry, CV4-7AL, UK
}}

\date{\today}
\maketitle

\begin{abstract}
We present experimental results for water wave turbulence excited by
piston-like programmed wavemakers in a water flume with
dimensions $6 \times 12 \times 1.5$ meters. Our main finding is that
for a wide range of excitation amplitudes the energy spectrum has a
power-law scaling, $E_\omega \sim \omega^{-\nu}$. These scalings
were achieved in up to one-decade wide frequency range,
 which is significantly wider than the
range available in field observations and in numerical simulations.
However, exponent $\nu$  appears to be non-universal. It depends on
the wavefield intensity and ranges from about 6.5 for weak forcing
to about 3.5 for large levels of wave excitations. We discuss our
results in the context of the key theoretical predictions, such as
Zakharov-Filonenko spectrum $\nu=-4$, Phillips spectrum $\nu = -5$,
Kuznetsov's revision of Phillips spectrum (leading to $\nu=-4$) and
Nazarenko's prediction  $\nu=-6$ for weak turbulence in finite
basins. We measured  Probability Density Function of the surface
elevation and good agreement with the Tayfun shape except values
near the maximum which we attribute to an anisotropy and
inhomogeneity caused by the finite flume size. We argue that the
wavenumber discreteness, due to the finite-size of the flume,
prevents four-wave resonant interactions. Therefore, statistical
evolution of the water surface in the laboratory is significantly
different than in the open ocean conditions.

\end{abstract}

\section{Introduction}

Understanding statistical properties of the water surface waves is
of great importance for a vast range of applications, from
navigation and industrial activities at sea to modelling the
transfer of momentum, heat, gases and aerosols through the air-sea
interface. The latter are an important factor in improving weather
and climate modelling and yet they are still very poorly understood.
For example, a number of specialized centres and regional agencies
provide a wave forecast for several days using certain assumptions
about the wave statistics. Besides wave forcing by the wind, the
other major ingredients of the models used by these centres are the
energy dissipation by wave breaking and the nonlinear interaction
that takes place among waves with different length.

This nonlinear interaction is described by a theory called wave or
weak turbulence theory (WT) introduced in the sixties by Zakharov
and Hasselman (see \cite{ZLF} for history and details of WT). WT
starts with the Euler equations for water with free surface in
gravity field and derives a kinetic equation (KE) for the wave
spectrum \cite{hasselmann}. The central prediction of WT is a steady
state energy spectrum obtained by Zakharov and Filonenko (ZF)
\cite{ZF},
\begin{equation}
E_\omega \propto \omega^{-\nu}
\label{power_law}
\end{equation}
with the index $\nu =4$. A number of hypotheses are needed in order
to derive KE: weak nonlinearity, random phase approximation,
homogeneity of the wave filed. Conditions of validity of these
hypotheses are still poorly understood. In addition, there exist an
alternative approach originated by Phillips which assumed that the
wave spectrum is not determined by weakly nonlinear dynamics but by
strongly nonlinear sharp-crested waves which arise due to
wavebreaking. This results in well-known Phillips power-law spectrum
(PH) \cite{phillips} with exponent  $\nu=  5$. Perhaps even stronger
uncertainty exists about the mechanisms and the form of wavebreaking
arising in the random wave field.

Thus, there has been a significant effort to study the random waves
and to test assumptions and predictions via field observations,
laboratory experiments, numerical modelling. There are advantages
and limitations in each of these approaches. The field observations
of waves are done using buoys or ships \cite{ships} or remotely from
aeroplanes \cite{remote}. They are extremely important because they
measure the processes directly as they occur in nature. However,
field observations are costly, have somewhat limited accuracy and
lack of control over the observation conditions. This makes
smaller-scale laboratory experiments and numerical modelling
valuable parts of research.

Numerical experiments are naturally cheaper and controllable, but
at present resolution they are not able to realise WT setup.
Indeed, for the WT approach to be relevant
it is essential that the nonlinear resonance broadening is
greater than the spacing between the neighbouring wavenumbers, -
otherwise most of the wave resonances are lost.
As estimated in \cite{sandpile}, this implies a condition
on the minimal angle of the surface elevation
\begin{equation}
\gamma > 1/(kL)^{1/4},
\label{alpha}
\end{equation}
where $L$ is the size of the basin.
This is a very severe restriction meaning, e.g., that
for a ten-kilometer wide gulf and meter-long waves
one should have $\gamma > 0.1$.

On the other hand,
Even for very small excitation amplitudes some resonances survive
\cite{clnp,meso,lnp} and it is possible that they can support the
energy cascade through scales even when condition (\ref{alpha}) is
not satisfied. Particularly, there have been numerous papers
reporting results of the direct numerical simulations and claiming
to confirm ZF spectrum, even though to satisfy condition
(\ref{alpha}) at a weakly nonlinear level  $\gamma \sim 0.1$ one
would have to compute at minimum $10000 \times 10000$ resolution
which is much greater than the resolution of all previous
simulations. Such a case, where the scalings and a qualitative
behavior are the same as in an infinite system, and yet there is a
quantitative disagreement (e.g. in the energy cascade rate) was
called in Ref \cite{meso} ``mesoscopic turbulence''.

On the other hand, due to necessity of numerical dissipation at high
wavenumbers (to avoid the bottleneck effect) the inertial interval
was very modest, - one decade in $k$ i.e. less than half-decade in
$\omega$ for gravity waves. This makes it very hard to judge about
the exact value of the slope, particularly deciding between the
$\omega^{-5}$ and the $\omega^{-4}$ spectra. Also, the water surface
equations used in computations are truncated at the level of cubic
nonlinearity and, therefore, become invalid for describing strongly
nonlinear events, particularly the wavebreaking.

In this situation, performing laboratory experiments realising
water wave turbulence looks especially attractive.
 Indeed, such experiments are significantly
cheaper and controllable than the field observations and yet they
are free form the artefacts of numerical
modelling due to truncation and artificial dissipation,
and they allow to resolve a significantly greater inertial
range of scales.

The present paper describes the results of one of such experiments.
Let us put our work in the context of the previous wavetank
experiments aimed at studying nonlinear dynamics of random
waves\footnote{Here we are not concerned with the short-time
experiments aimed at testing scaled models of sea installations or
ships where the spectrum was pre-conditioned by wavemakers and the
waves life time is not long enough for the nonlinearity to evolve.}
Our work differs from the previous experiments in two respects. The
wavetank is larger and has significant sizes in both horizontal
directions that provides and opportunity to observe 2D wave
evolution. We use a piston-type wavemaker rather than wind to
produce waves. The wavemakers are more suitable for testing WT
theory because they can force at low frequencies only while the wind
forcing is spread over the wide frequency range. As a result,
wavemakers enable to leave a more pure inertial range of scales for
with ZF prediction was made. Larger size is important for minimising
the finite-size effects and the $k$-space discreteness. Yet, as we
will report in the present paper, the size of our tank is not large
enough for WT theory to be applicable, and the wave field statistics
was strongly affected by the finite-size effects.

\section{Theoretical background}

Below we describe the theoretical background and predictions that
will serve us as reference points in interpretation of our
experimental results. First of all, let us consider the predictions
for the wave energy spectrum which is defined as,
\begin{equation}
E_\omega = \int e^{i \omega t'} \langle \eta({\bf x}, t)
\eta({\bf x}, t+t') \rangle \, dt',
\end{equation}
where $\eta({\bf x}, t)$ is the surface elevation at time $t$ and
location in the horizontal plane ${\bf x}$. Here, the the integral
is taken over a time window, and  angle brackets mean ensemble
averaging over realisations (a number of chosen time windows in the
entire signal record). For statistically steady and homogeneous
state, $E_\omega$ is independent of $t$ and ${\bf x}$.

\subsection{Weak turbulence theory, ZF spectrum.}

As we already mentioned in the introduction, WT theory considers
weakly nonlinear random-phase waves in an infinite box limit. The
central object of WT is KE for the wave spectrum, which in the
context of water gravity waves called Hasselmann equation
\cite{hasselmann}. This equation is quite lengthy and for our
purposes it suffices to say that ZF energy spectrum
$$E_\omega \propto \omega^{-4}$$
 is an exact solution
of Hasselmann equation which describes a steady state
with energy cascading through an inertial range of scales
from large scales where it is produced to the small scales
where it is dissipated by wavebreaking.

It is crucially important that in deriving KE, the limit of an
infinite box is taken before the limit of small nonlinearity. This
means that in a however large but finite box, the wave intensity
should be strong enough so that the nonlinear resonance broadening
is much greater than the spacing of the $k$-grid (corresponing to
Fourier modes in finite rectangular box). This is precisely the
condition which, due to lack of available resolution, has never been
satisfied in numerical experiments. As we will see below, this is
also the reason why we never see the WT regime in our laboratory
experiment.

\subsection{Phillips spectrum and its relatives.}

An easiest way to derive the PH spectrum is to assume that the
gravity constant $g$ is the only relevant  dimensional physical
quantity. Then the wave energy spectrum is uniquely determined in
terms of $g$ and $\omega$ based on the dimensional analysis
\cite{phillips},
\begin{equation}
E_\omega = g^{2} \omega^{-5}.
\label{ph}
\end{equation}
It is quite clear that this argument is equivalent to saying that
the linear term is of the same order as the nonlinear one in the
water surface equations in Fourier space. Such a balance of linear
dispersion  and nonlinear terms is typical for soliton-like
nonlinear structures.

Physically, the PH spectrum is usually associated with the sharp
crested waves, so that the short-wave Fourier asymptotics are
dominated by discontinuous slopes of such wavecrests. Assuming
first, that such a discontinuity is happening at an isolated point
(i.e. in a cone-like structure) we get for the one-dimensional
energy spectrum in wavenumber space
\begin{equation}
E_k \propto k^{-3}.
\label{ph_k}
\end{equation}
Second, assuming that transition from the  $k$-space
to the $\omega$-space should be done according
to the linear wave relation $\omega = \sqrt{gk}$, we
arrive at the PH spectrum (\ref{ph}).

Kuznetsov \cite{kuznetsov} questioned both of these assumptions and
argued that (i) slope break occurs on one-dimensional lines/ridges
rather than zero-dimensional point/peaks, and (ii) that the
wave-crest is propagating with preserved shape, i.e. $\omega \propto
k$ should be used instead of the linear wave relation $\omega =
\sqrt{gk}$. This assumptions give $E_\omega \propto \omega^{-4}$,
i.e. formally the same scaling as ZF, even though the physics behind
it is completely different.

Finally, it was proposed in Ref \cite{cnn} that wavecrest ridges
may have non-integer fractal dimension somewhere in the range $0<D<2$.
Assuming, following Kuznetsov,  $\omega \propto k$, we
have in this case
\begin{equation}
E_\omega \propto \omega^{-3 - D}.
\label{fractal}
\end{equation}

In our experimental results reported below, at large forcing levels
we observe the prominent wavebreaking events the role of which in
forming spectra is quite apparent. However, the spectrum exponent
appears to be dependent on the forcing intensity. Possibly this
could be due to dependence of   wavebreaking morphology and
dimension $D$ on the wave turbulence intensity.

\subsection{Discrete  wave turbulence.}

Here, we will briefly describe the theory suggested in
\cite{sandpile} for the case of very weak turbulence in discrete
$k$-space. As we already mentioned, for the WT mechanisms to work,
the four-wave resonances must be broad enough to cover
simultaneously many discrete $k$-modes, and this condition in terms
of the surface slope $\gamma$ gives the estimate (\ref{alpha}). What
happens if the waves are so weak that this condition is not
satisfied? In this case, the number of exact and quasi four-wave
resonances will be drastically depleted
\cite{kartashova1,kartashova2,sandpile}. This will lead to the
arrest of the energy cascade from long to short waves and,
therefore, there will be an accumulation of the spectrum near the
forcing scale. Such an accumulation will proceed until the intensity
is strong enough for the nonlinear broadening to become comparable
to the $k$-lattice spacing, i.e. when condition (\ref{alpha}) will
become marginally satisfied. At this point, the four-wave resonances
will get engaged and the spectrum ``sandpile'' will tip over toward
the higher wavenumbers. This process will proceed until the whole
$k$-space will be filled by the spectrum having a critical slope
determined by the condition that the wave-resonance broadening is of
the order of the $k$-grid spacing for all modes in the inertial
range. This condition gives the following spectrum
$$E_\omega \propto \omega^{-6}.$$

\subsection{Probability density functions.}

Some important information about the wave field statistics, not
contained in the spectra, can be accessed by measuring the
probability density function (PDF) of the surface elevation. We
remind for reference that homogeneous isotropic wave fields with
random independent phases of all of its modes are characterized by
Gaussian PDF shape. The Gaussian shape is expected for linear and
weakly nonlinear waves, provided the conditions of homogeneity and
isotropy are satisfied. The PDF for stronger nonlinearities
 was obtained by Tayfun
\cite{tayfun} using a model where the wave field is made of
independent  weakly nonlinear Stokes waves whose
first harmonics are gaussian. Tayfun distributions where found to be
in good agreement with numerical simulations with wide-angle
quasi-isotropic wavefields \cite{dysthe} and to much lesser extent
in narrow-angle distributions \cite{onorato_exp, onorato_nontayfun}.

Another interesting statistical object is a PDF of the spectral
intensities (squared modules of the Fourier coefficients) at  a
particular fixed frequency $\omega$. Based on a generalized WT
approach, it was found theoretically in Ref. \cite{clnp} that such a
PDF may differ from the Rayleigh distribution (corresponding to
Gaussian wave fields) by significantly fatter tails. This
corresponds to a probability to observe strong waves more frequently
than for Gaussian waves (a ``freak wave'' effect).

\begin{figure}
\includegraphics[width=110mm,height=80mm]{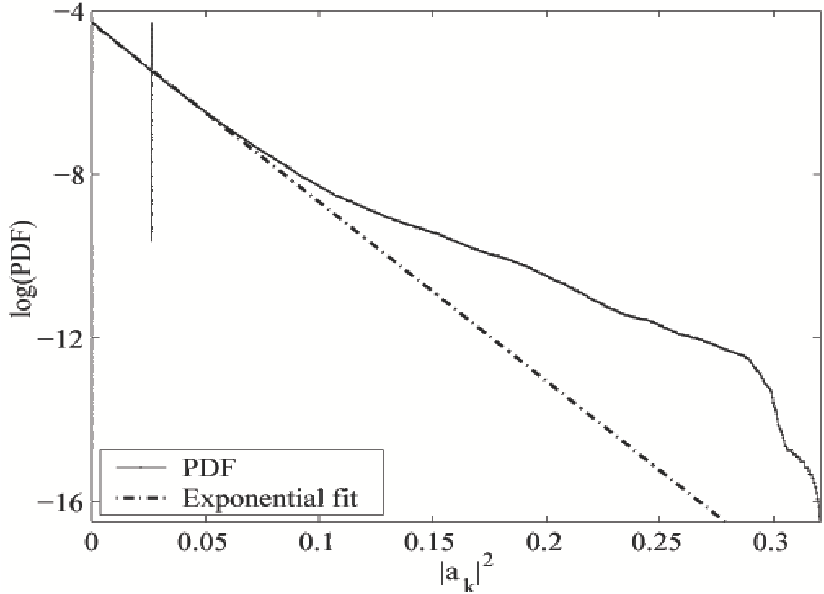}
\caption{Numerically obtained PDF of spectral intensity in a small
band around $\omega$ equal to the eight minimal wave frequencies
\cite{clnp,lnp}.} \label{num_pdf}
\end{figure}

A typical form of such a  PDF obtained in numerical experiments
\cite{clnp,lnp} by applying band-pass filtering with $\Delta \omega
\ll \omega$  is shown in Figure \ref{num_pdf}. Indeed, we see a
significant deviation in the tail from Rayleigh shape (straight line
in Figure \ref{num_pdf}). Later in the present paper we will observe
a similar effect in our experimental results.

\section{Experimental setup}

\begin{figure}
\includegraphics[width=140mm,height=80mm]{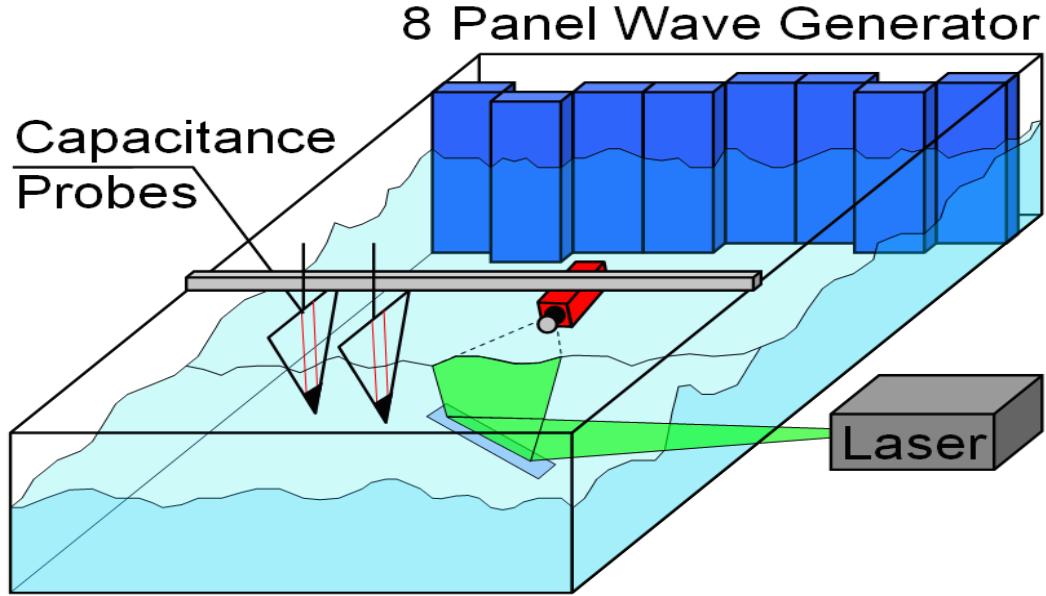}
\caption{Experimental setup at ``The Deep'' flume.}\label{deep}
\end{figure}

Our  experiments with surface gravity waves were conducted in a
rectangular tank with dimensions 12 x 6 x 1.5 meters filled with
water up to the depth of 0.9 meters. The wavemaker consists of 8
vertical paddles of width 0.75 m covering the full span of one short
side of the tank, see Figure  \ref{deep}. Each paddle can oscillate
horizontally in the direction perpendicular to its face plane.
Amplitude, frequency and phase can be set for each panel
independently allowing, in principle, to control directional
distribution of the generated waves. A motion controller is used to
program parameters of the generated wavefield by specifying
amplitude, frequency distribution and a number of wavevector
directions.

\begin{figure}
\includegraphics[width=165mm,height=50mm]{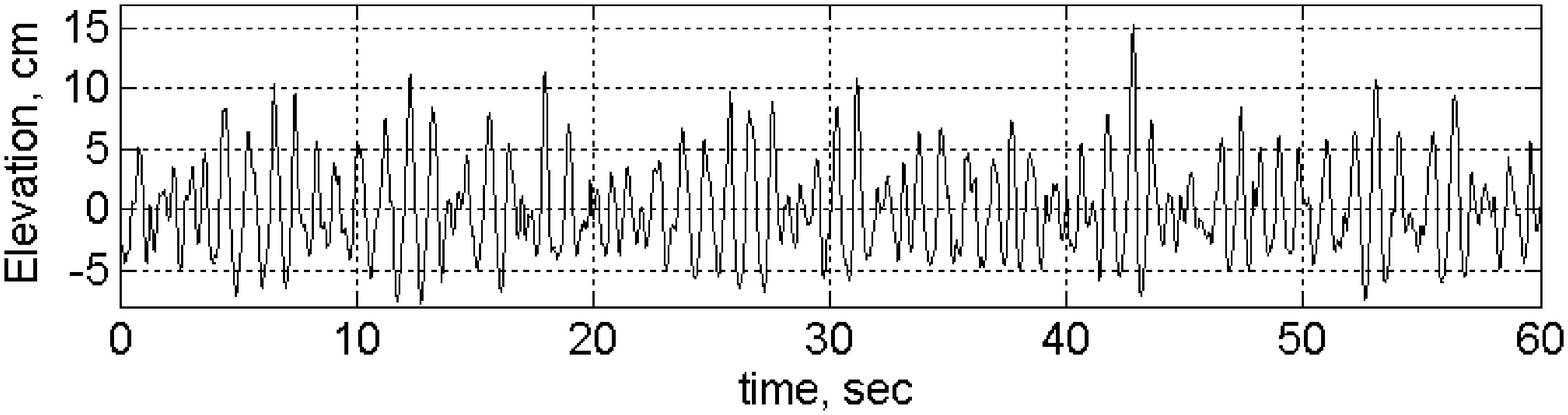}
\caption{A typical signal for the time evolution of the surface
height at one of the gauges.} \label{signal}
\end{figure}

In all experiments described in the present paper, the frequency
bandwidth generated by the wavemaker  had a fixed broadband
distribution in the range 0.4 to 1.2 Hz, which corresponds to the
wavelengths  1 to 3.9 m. To get more spatially isotropic
excitations, the wavevector directions  were chosen within a
90-degree sector enabling the multiple reflections  even for the
shortest excited wavelengths. The main control parameter was an
excitation amplitude on the wavemaker, and we performed the
experiments at different values of this parameter to study
dependence of the spectrum and PDFs on the average wave turbulence
intensity. The surface elevation was measured simultaneously by two
capacitance gauges, A and B, positioned in the middle part of the
tank. A-gauge was fixed while another could move parallel to the
short flume side, so that the distance between the gauges could vary
from 0.1 to 2.5 m, see Figure \ref{deep}. In most of the reported
experiments this distance was 40 cm. Each probe consists of two
parallel vertical wires separated by 1 cm distance. Due to the
difference in the dielectric permeability of water and air, the
coupling capacitance between the wires depends on their submergence
depth. This capacitance was measured using sinusoidal AC with
different frequencies used for each gauge (60 and 90 kHz) to avoid a
crosstalk between them. Signals from the gauges were amplified by
two lock-in amplifiers. The outputs from the amplifiers were
digitized by a multi-functional board  (NI6035, National
Instruments) and stored as a waveform using LabView software.
Typical parameters of the acquired signals were as follows. The
bandwidth at the lock-in amplifier output was 32 Hz, the sampling
rate was 400 Hz for each channel, the minimum acquisition time was
2000 seconds. The gauges were calibrated before the measurements in
the same tank with a stationary water surface. The experimental data
set contains a number of two-channel waveforms acquired at different
amplitudes of the generated waves and for two different
configurations of the wavemaker paddles. In one configuration all
eight paddles were used, in others only seven or six paddles worked
while one or two corner paddles were disconnected. Although using
the six and seven paddle configurations were caused by technical
reasons, these additional data enable us to compare results with
different wave excitation geometries. The measurement procedure was
the same during the whole experiment and consisted of setting the
excitation wave amplitude, waiting for 15 minutes of transient time,
measuring and writing the signals during the next 30-60 minutes. A
typical time signal, the wave elevation vs time, is shown in Figure
\ref{signal}

Collected waveforms were processed using Matlab. The common initial
part for all data processing procedures included high-pass filtering
with the time constant 0.01 s eliminating a slow drift of signals,
decimation to the sampling frequency 50 Hz, and elimination of the
initial transitional part of signals. At the next stage of
processing we evaluated the statistical characteristics of  waves:
the spectra (by applying the fast Fourier transform), the PDF of the
surface height and the PDF of Fourier modes (by band-pass
filtering).

One of the quantities we have used to characterize  the wave field strength
was the RMS of the wave height,
\begin{equation}
A= \sqrt{\langle (\eta - \eta_0)^2 \rangle},
\label{A}
\end{equation}
where $\eta = \eta(t)$ is the water surface elevation at the gauge
position and  $\eta_0$ is its mean value. Here, the angle brackets
denote the time averaging over the whole measuring interval. As a
characteristics of nonlinearity, we used the mean slope at the
energy containing scale,
\begin{equation}
\gamma = k_m A,
\label{gamma}
\end{equation}
where  $k_m$ is the wavenumber corresponding the the maximum of the
energy spectrum. In all our experiments $k_m$ was approximately the
same and located in the forcing range, $k_m \approx 4.0 $m$^{-1}$
which corresponds to the wavelength $\lambda \approx 1.6$ m. Note
that in the same experiment waves with different frequencies usually
have different nonlinearities. Depending on the spectrum slope
$\nu$, the nonlinearity may be smaller at higher frequencies (for
$\nu >5$) or greater (for $\nu <5$). One can see that PH spectrum is
a borderline case in which the nonlinearity is the same at all
frequencies.

In our experiments we covered the range of excitations from very
weak waves with mostly smooth surface and occasional seldom
wavebreaking, $A \approx 1.3$ cm and $\gamma \approx 0.05$, to very
strong wave amplitudes characterized by a choppy surface with
numerous wavebraking events, $A \approx 5.2 $ cm and $\gamma \approx
0.21$.

\subsection{Spectra}

\begin{figure}
\includegraphics[width=110mm,height=80mm]{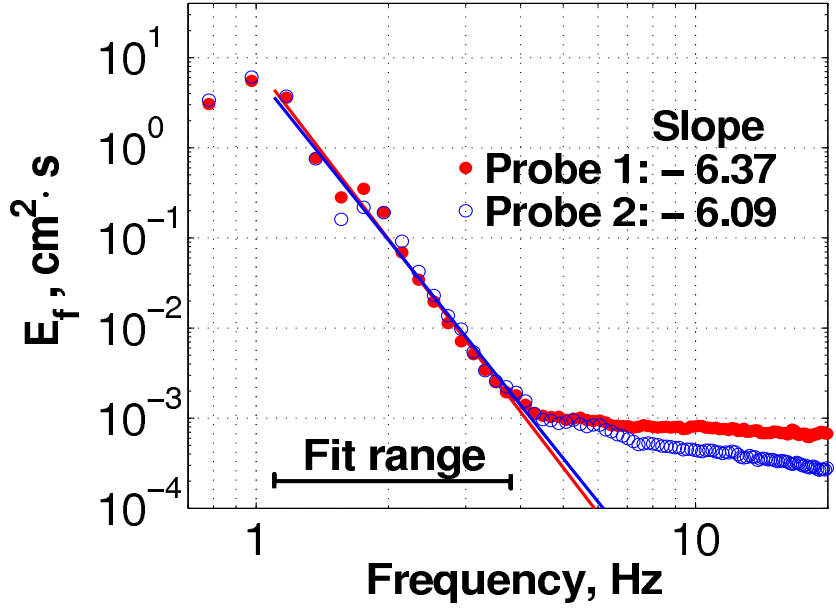}
\caption{Spectrum obtained in the 6-paddle experiment with low wave
intensity, $A \approx 1.9$ cm ($\gamma \approx 0.0.074$).}
\label{spec_weak}
\end{figure}

\begin{figure}
\includegraphics[width=110mm,height=80mm]{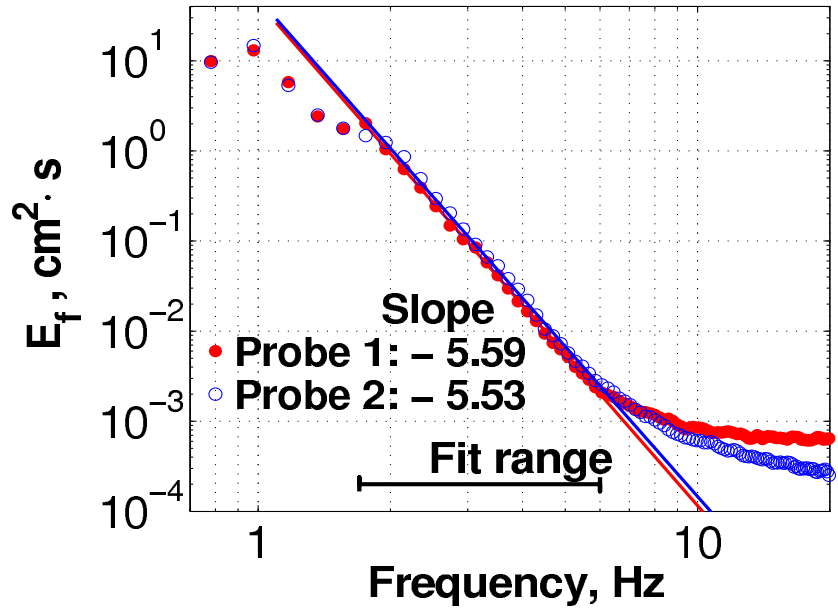}
\caption{Spectrum obtained in the 6-paddle experiment with medium
wave intensity, $A \approx 3.05$ cm ($\gamma \approx 0.12$).}
\label{spec_medium}
\end{figure}

\begin{figure}
\includegraphics[width=110mm,height=80mm]{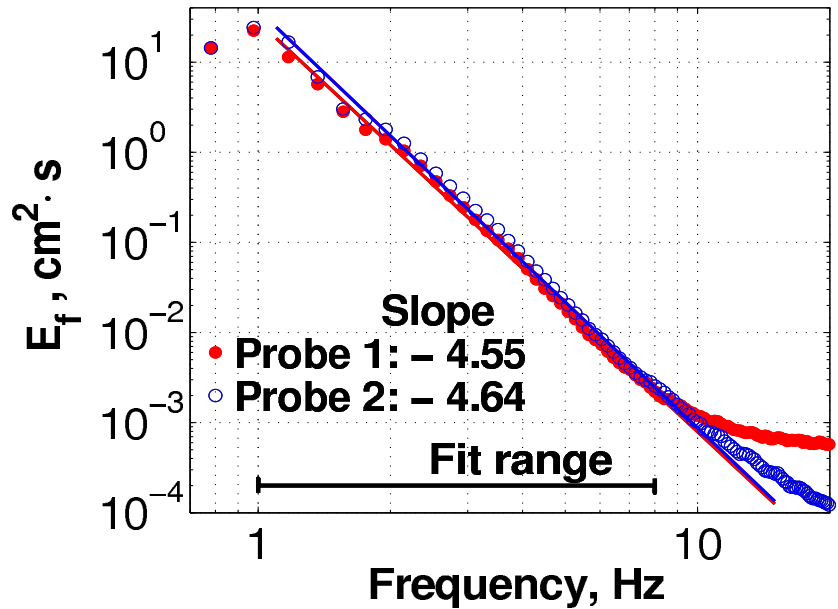}
\caption{Spectrum obtained in the 7-paddle experiment with high wave
intensity, $A \approx 3.73$ cm ($\gamma \approx 0.15$).}
\label{spec_strong}
\end{figure}

To find spectra we used the Welch algorithm with the Hanning window
of length 256 points (5.12 s) and the averaging performed over about
1000 spectral estimates for each signal record. Typical results for
the spectra for several different levels of nonlinearity is shown in
Figures \ref{spec_weak}, \ref{spec_medium} and \ref{spec_strong}.

\begin{figure}
\includegraphics[width=110mm,height=80mm]{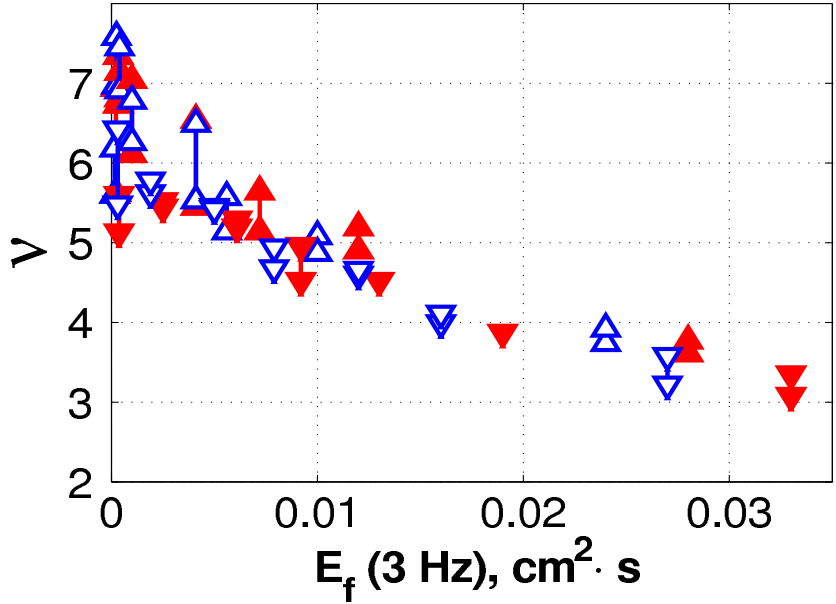}
\caption{Plot of spectral slopes $\nu$ in the 6-paddle and 7-paddle
experiments  versus the wave intensity at 3 Hz, see text for
details.} \label{slopes}
\end{figure}

Horizontal lines show  an interval where the spectrum
slope (index) is estimated using linear least-square logarithmic fit.
The maxima of  spectra in all cases are at about 1.2 Hz which
 corresponds to the highest
frequency in the pumping wave. Lower frequencies have less power relative
to the spectrum maximum due to the dumping of long waves by the bottom friction.
One can see that the scaling behavior is not-so-well formed in
Figure \ref{spec_weak} which corresponds to the weakest wave field
intensity $A \approx 1.85$ cm and $\gamma \approx 0.074$, and this
leads to some
 uncertainty in the value of the slope and its sensitivity to the
choice of the fitting range. On the other hand, the spectra
corresponding to the medium and the strongest intensities, shown in
Figures  \ref{spec_medium} and \ref{spec_strong} respectively,
exhibit clear scaling ranges which are up to one decade wide and
have well-defined slopes. The results for the slopes obtained in
different experiments, including the cases with six and seven
working paddles, are summarized in Figure \ref{slopes} with the
slope uncertainty indicated by the vertical bars. The bars were
constructed by varying the fitting range and finding the minimal and
the maximal slopes. The $x$-axis represents the intensity value of
the spectrum at 3Hz frequency, $E_{3Hz}$. The reason for this choice
of measure of the wave turbulence intensity is that this quantity
appears to be more universal than the RMS of the surface elevation,
$A$, or  RMS of the slope parameter $\gamma$. This is because the
latter two are dominated by the energy containing scale which
appears to be different for the 6-paddle and the 7-paddle series of
experiments. On the other hand, $3$ Hz is within the scaling
(equilibrium) range of all spectrum data sets, and the data points
for the slope values $\nu$   for the six- and seven-paddle
experiments collapse much better on the same curve if $E_{3Hz}$ is
used as a measure of intensity.

The general trend, clear in  Figure \ref{slopes}, is that the slope
$\nu$ is much steeper for the weak wave fields with respect to the
strong ones. One can see that at small amplitudes the data scatter
and uncertainty are much greater at low intensities than for
stronger wave turbulence. For experiments with the minimal
intensity, $A \approx 1.85$ cm and $\gamma \approx 0.074$, we have
$\nu \approx 6.3 \pm 1.2$, which is in good qualitative agreement
with the prediction $\nu = 6$ made in \cite{sandpile} for the
critical spectrum where the nonlinear resonance broadening is of the
same magnitude as the mean spacing between the discrete $k$-modes.
Thus, we indirectly confirm that the $k$-space discreteness (caused
by the finite flume size) plays a defining role in shaping the
frequency spectrum at low wave excitations. We see that the
condition (\ref{alpha}) is not quite satisfied,
$$
 \gamma \approx  0.074 <  1 /(k_m L)^{1/4} \approx 0.4.
$$
Note that  (\ref{alpha}) is only an order of magnitude estimate so a
factor of 5 discrepancy could easily be an order one coefficient
(like $\pi$)
 unaccounted by (\ref{alpha}).
On the other hand, it might also mean that an efficient
energy cascade may start even at the level or
resonance broadening which
remains less than the $k$-grid spacing.
In this case not all of the four-wave resonances are equally
engaged, and that most of the energy cascades from low to high wavenumbers
are carried by most active quartets of  modes  which
causes extra anisotropy of wave turbulence.
Such an anisotropy could also be a natural reason for deviations from
 the pure power-law scaling seen as a significant slope
uncertainty and scatter at low intensities in Figure \ref{slopes}.

At large wave field intensities, one can see   in Figure
\ref{slopes} much better scaling behavior with significantly smaller
scatter or uncertainty in the slope values. There is a range of
intensities where the PH slope $\nu =5$ is observed, and we report
that wave breaking events were common for such intensities. At
higher intensities, one can see the  $\nu=4$ slope which is
predicted by both ZF and Kuznetsov theories \cite{ZF,kuznetsov}.
 However, the water surface was
visibly very choppy with numerous frequent wavebreaking and high
values of the surface slope, $\gamma >0.15$, and rules out the weak
nonlinearity assumption which is the basis of ZF theory \cite{ZF}.
Kuznetsov theory \cite{kuznetsov} is more likely to be relevant to
these conditions, because it derives $\nu =4$ value from considering
strongly nonlinear  wavecrests with sharp 1D ridges
 and the speed of which is nearly constant while
they pass the height gauge. However, there is no visible plateau in
Figure \ref{slopes} at  $\nu =4$ value and  $\nu $ takes bigger
values at lower intensities and lower values for greater amplitudes
(reaching  $\nu =3.3$ for the maximum intensity of $\gamma \approx
0.21$). As we mentioned before, such a change of slope could be
explained by changing the fractal dimension of the wavecrest ridges,
from $D=0$ cone-like splashes giving $\nu =3$ (this is not PH
because $\omega \sim k$ is assumed instead of $\omega \sim
\sqrt{k}$) at smaller amplitudes,
 set of $1D$ lines giving  $\nu=4 $ at larger amplitudes
(Kuznetsov) to more complex fractal curves  at lower
intensities with $1<D<2$ giving $4<\nu <5$. Note that Ref.
\cite{cnn} obtained value $D=3/2$ by considering scalings of the
higher moments within the wave turbulence formalism.

\subsection{Probability density functions.}

\begin{figure}
\includegraphics[width=110mm,height=80mm]{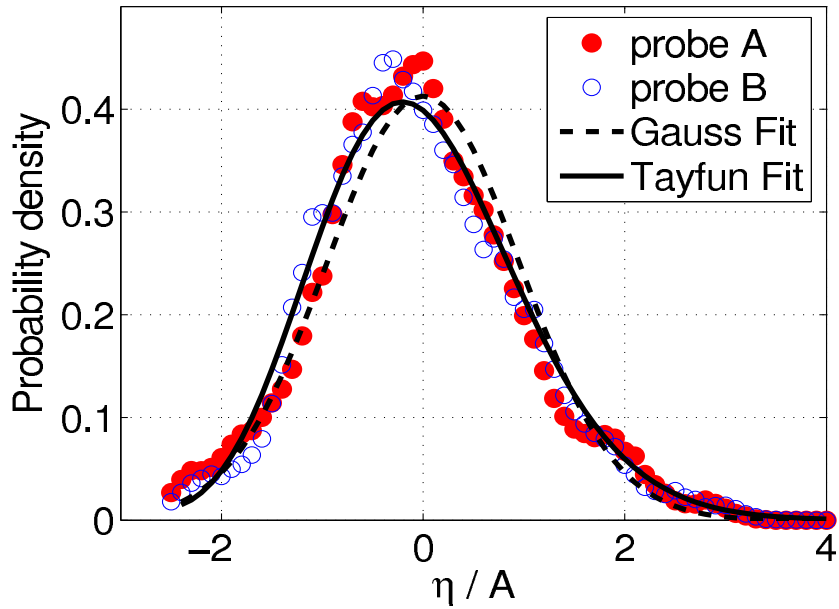}
\caption{Normalized PDF of the surface height at locations of gauges
A (dots) and B (triangles)  in the 6-paddle experiment for mean
intensity $A=1.85 cm$ ($\gamma=0.074$).} \label{pdf1}
\end{figure}

\begin{figure}
\includegraphics[width=110mm,height=80mm]{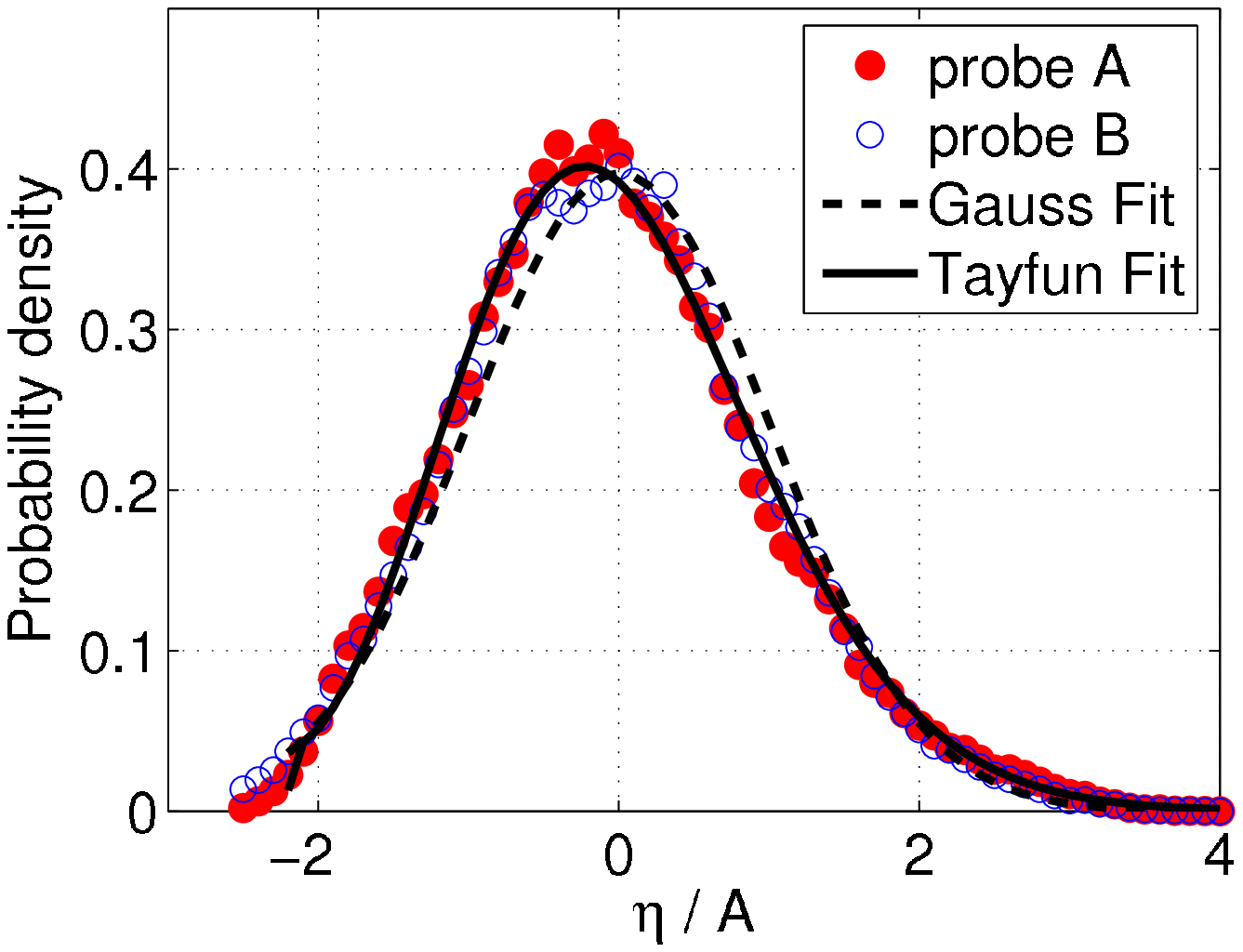}
\caption{PDF of the surface height at locations of gauges A (dots)
and B (triangles)  in the 6-paddle experiment for mean intensity
$A=3.05 cm$ ($\gamma=0.12$).} \label{pdf2}
\end{figure}

\begin{figure}
\includegraphics[width=110mm,height=80mm]{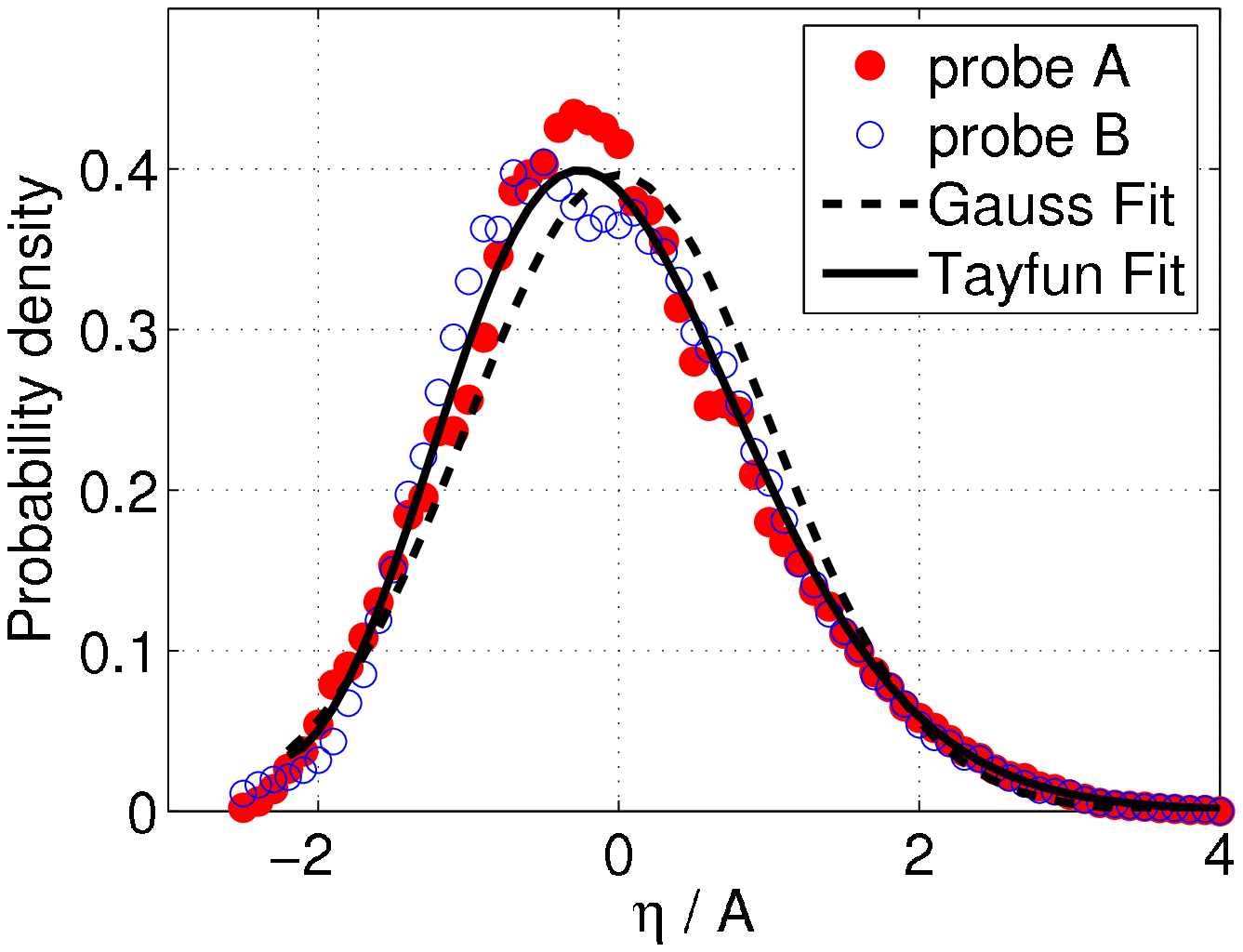}
\caption{PDF of the surface height at locations of gauges A (dots)
and B (triangles)  in the 7-paddle experiment for mean intensity
$A=3.73 cm$ ($\gamma=0.15$).} \label{pdf3}
\end{figure}

Our results for PDF of the surface elevation measured for different
wave intensity levels, for different numbers of working paddles (6
or 7) and for different locations (at gauges A and B) are shown in
Figures \ref{pdf1}, \ref{pdf2} and  \ref{pdf3}. One can see good
qualitative agreement of PDFs with the Tayfun distribution at all
intensities, - low at Figure \ref{pdf1}, medium at Figure \ref{pdf2}
and high at Figure \ref{pdf3}. However, one can notice rather
irregular deviations, especially near the PDF maximum corresponding
to probability of waves with  less-than-mean intensities. We have
checked that these deviations are real and are not fluctuations  due
to  insufficient statistical   data. To do this, we split a long
record corresponding to one of experiments into two shorter
sub-sets, and we observed that the resulting PDFs for both sub-sets
were identical to the PDF obtained from the original long record.
One can also see that the observed deviations from Tayfun were
different at the same experiment for two different locations (i.e.
at gauges A and B).

\begin{figure}
\includegraphics[width=110mm,height=80mm]{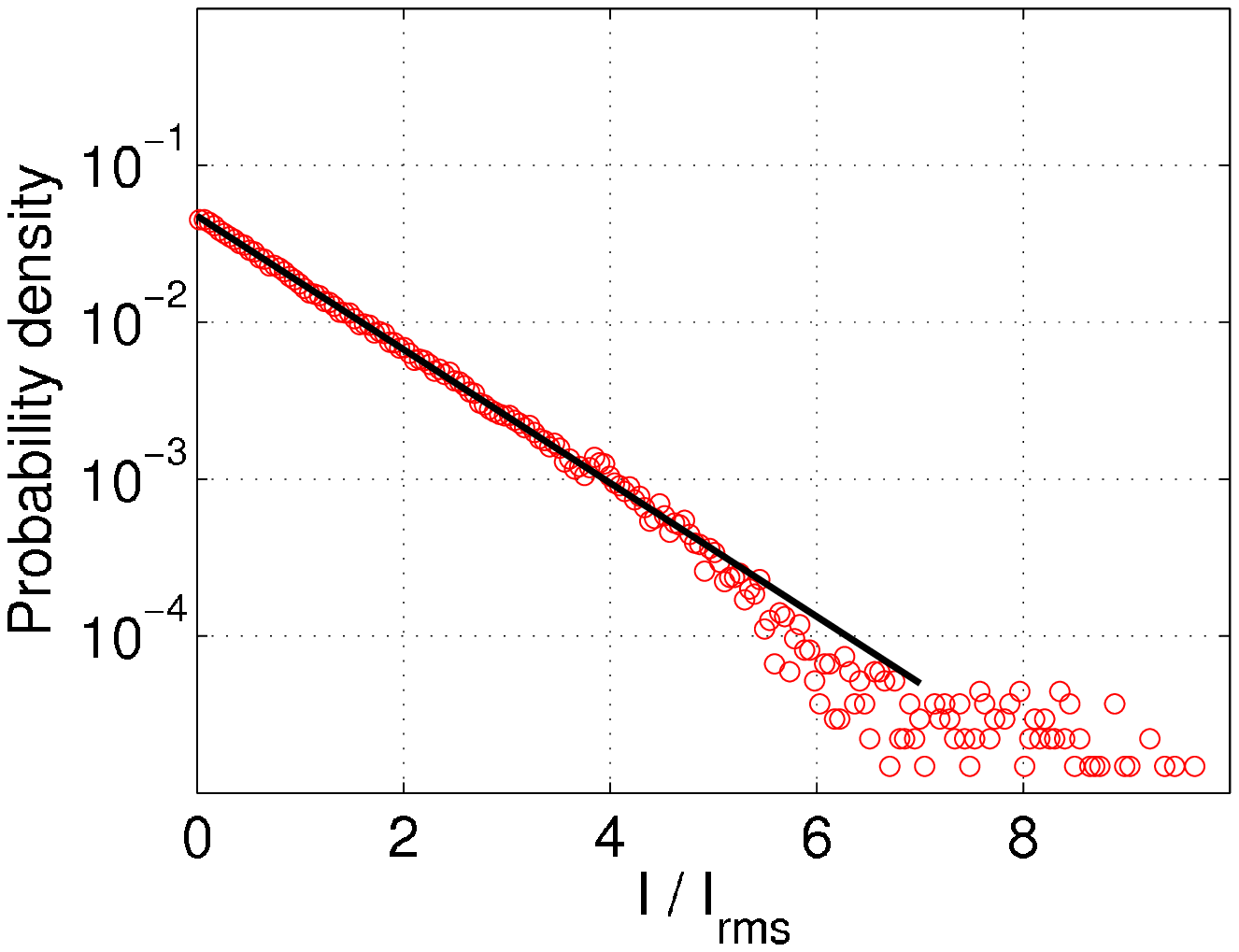}
\caption{PDF of the spectral intensity band-pass filtered with a
frequency window $\Delta f =1 Hz$ centered at $f =6 Hz$. The signal
corresponds to probe B in a low-intensity experiment with mean
elevation $A=1.85 cm$ ($\gamma=0.074$)} \label{kpdf1}
\end{figure}

\begin{figure}
\includegraphics[width=110mm,height=80mm]{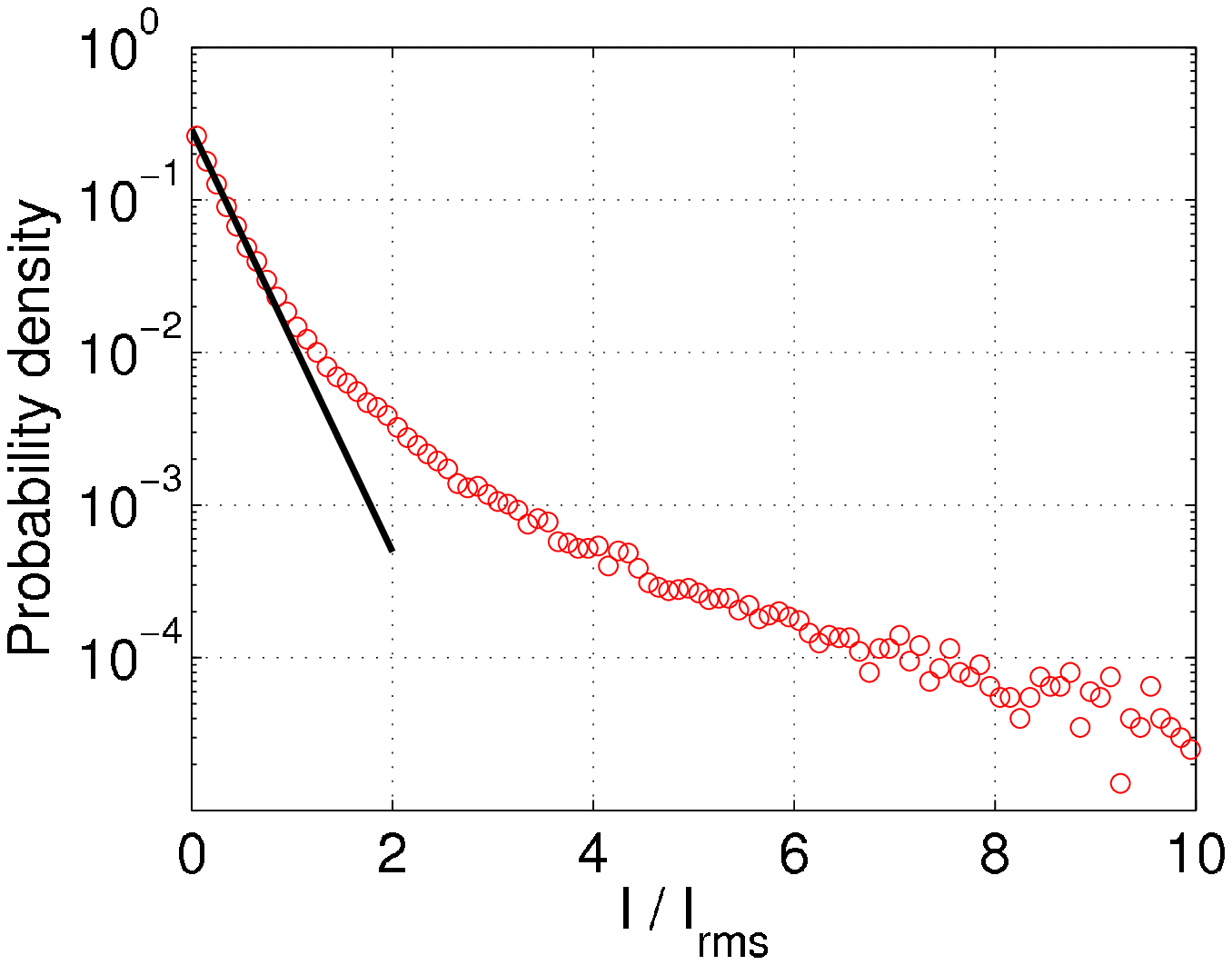}
\caption{PDF of the spectral intensity band-pass filtered with a
frequency window $\Delta f =1 Hz$ centered at $f =6 Hz$. The signal
corresponds to probe B in a high-intensity experiment with mean
elevation $A=3.73 cm$ ($\gamma=0.15$)} \label{kpdf2}
\end{figure}

Now let us consider PDFs of the spectral intensities obtained by
band-pass filtering of the time signal with a narrow pass window
$\Delta \omega$ around a particular frequency $\omega$ ($\Delta
\omega \ll \omega$). These PDFs for the experiments with low and
high mean intensities of the wave field are shown in Figures
\ref{kpdf1} and \ref{kpdf2} correspondingly. One can see a similar
picture as in numerical results shown in Figure \ref{num_pdf},
namely much higher with respect to the Rayleigh distribution
probability of strong waves. This is a nonlinear effect and,
therefore, it is natural that it is more pronounced for higher
intensities (i.e. in Figure \ref{kpdf2}).

\section{Conclusions}

One of the main observations derived from our experimental results
is that the slope of the wave spectrum is not universal:  it takes
high values, around 6.5, for weak wave fields and gradually
decreases as one increases the wave field intensity to the value of
about 3.5 when the wave field is very choppy with a lot of
wavebreaking. In the absence of forcing, one would expect KAM-like
behavior for most of the $k$-space because only a tiny fraction of
modes on the discrete $k$-latice can satisfy the exact four-wave
resonant conditions \cite{kartashova1,kartashova2,sandpile}.
However, our system had a continuous forcing and the spectrum can
grow until it reaches a critical slope where the nonlinear resonance
broadening becomes comparable to the $k$-grid spacing, and the
spectrum ``sandpile'' can ``tip over'' producing intermittent
avalanches carrying energy from large to small scales. Such a
critical spectrum was argued in Ref \cite{sandpile} to have slope 6,
which is in a good qualitative agreement with our observations. For
larger amplitudes, we see a gradual decrease of the slope, possibly
due to the sharp wavecrests whose fractal dimension decreases with
increasing wave intensity (from $D=0$ for strongest and most choppy
waves, to $D=1$ for
Kuznetsov spectrum, to $1<D<2$ for weaker
wave fields. We emphasize that at this point dependence of $D$ on
the wave intensity is purely speculative, and a careful study of the
wavebreaking morphology is needed to be done to measure $D$ in
experiments.

Another aspect of our study was the measurement of PDFs of the wave
elevations which, as expected, agree well with the Tayfun
distribution even though the flume finite-size effects are indeed
seen to lead in irregular deviations from the perfect Tayfun shape.
We also observed higher than in Gaussian fields probability of
strong wave modes in the way similar to theoretical predictions and
numerical simulations of \cite{clnp,lnp}.

The major feature of our experiments was that the WT regime was
never achieved: with increasing wave intensity the nonlinearity
becomes strong before the system looses sensitivity to the $k$-space
discreteness.  Particularly, most of the four-wave resonances
remained arrested leading to a depletion of the energy cascade from
low to high wavenumbers. Thus, being interesting in their own way,
laboratory waves undergo a significantly different statistical
evolution from the one of their open-sea counterpart. In our future
work we will explore possibilities to ``confuse'' the laboratory
flume and make it ``forget'' about the finite-size effects via
breaking the regular $k$-space structure by distorting the
perfect-rectangle shape of the flume boundaries. The goal of such
exercise would be to increase realism of laboratory modeling of the
ocean surface processes.

\end{document}